\documentclass [aps,prb,twocolumn,superscriptaddress]{revtex4}
\usepackage{graphicx}
\usepackage{amsmath}
\usepackage{braket}
\usepackage{xcolor}
\usepackage{bm}
\usepackage{physics}
\usepackage{mathtools}
\usepackage{hyperref} 
\hypersetup{
	colorlinks   = true, 
	urlcolor     = black, 
	linkcolor    = black, 
	citecolor   = black, 
}

\begin{document}
\title{Strong Spin-Orbit Torque effect on magnetic defects due to topological surface state electrons in Bi$_{2}$Te$_{3}$}
\author{Adamantia Kosma}
\affiliation{Section of Condensed Matter Physics, Department of Physics, National and Kapodistrian University of Athens, Panepistimioupolis 15784 Athens, Greece}
\author{Philipp R\"u\ss mann}
\affiliation{Peter Gr\"unberg Institut and Institute for Advanced Simulation, Forschungszentrum J\"ulich and JARA, 52425 J\"ulich, Germany}   
\author{Stefan Bl\"ugel}
\affiliation{Peter Gr\"unberg Institut and Institute for Advanced Simulation, Forschungszentrum J\"ulich and JARA, 52425 J\"ulich, Germany}   
\author{Phivos Mavropoulos}
\affiliation{Section of Condensed Matter Physics, Department of Physics, National and Kapodistrian University of Athens, Panepistimioupolis 15784 Athens, Greece}
    
\begin{abstract}
We investigate the spin-orbit torque exerted on the magnetic moments of the transition-metal impurities Cr, Mn, Fe and Co, embedded in the surface of the topological insulator Bi$_{2}$Te$ _{3} $, in response to an electric field and a consequent electrical current flow in the surface. The multiple scattering problem of electrons off impurity atoms is solved by first-principles calculations within the full-potential relativistic Korringa-Kohn-Rostoker (KKR) Green function method, while the spin-orbit torque calculations are carried out by combining the KKR method with the semiclassical Boltzmann transport equation. We analyze the correlation of the spin-orbit torque to the spin accumulation and spin flux in the defects.  We compare the torque on different magnetic impurities and unveil the effect of resonant scattering. In addition, we calculate the resistivity and the Joule heat as a function of the torque in these systems. We predict that the Mn/Bi$_{2}$Te$_{3}$ is optimal among the studied systems.
\end{abstract}	

\maketitle

\section{\label{sc:intro}Introduction}
The field of spintronics,~\cite{zutic_spintronics:_2004,wolf_spintronics:_2001} that aims at controlling the electron spin degree of freedom, has proven to be a powerful tool to design devices with applications for information technology. A very active area in this field is the ``electrically controlled spintronics'' that is related to the manipulation of the magnetization by means of an electric field, allowing for high density of magnetic memory components in memory devices.~\cite{bhatti_spintronics_2017,chappert_emergence_2007}

The main research direction in this area concerns the current-induced spin torque effect that was pioneered by Slonczewski~\cite{slonczewski_current-driven_1996} and Berger~\cite{berger_emission_1996} in 1996. They introduced the concept of the spin transfer torque,~\cite{ralph_spin_2008} according to which a spin polarized current, emitted from a ferromagnetic layer which acts as the polarizer, causes a precession of the magnetization of a second ferromagnetic layer. This effect can be used for an electric-field control of Magnetoresistive Random-Access Memories (MRAMs)~\cite{apalkov_magnetoresistive_2016}, interpreting the ``up" or ``down" direction of magnetization as the logical states of a magnetic memory bit and accordingly writing the magnetic information. In the last few years, another type of current-induced spin torque, the spin-orbit torque (SOT),~\cite{manchon_current-induced_2019,manchon_theory_2008} has gained ground. Its main advantage is that the charge current is converted to a spin current~\cite{gambardella_current-induced_2011} allowing the control of magnetic states without the need of a polarizer. The SOT effect has been investigated mainly in ferromagnetic bilayers or multilayers theoretically~\cite{garate_influence_2009,freimuth_spin-orbit_2014-1,freimuth_direct_2015,geranton_spin-orbit_2015,geranton_spin-orbit_2016,mahfouzi_first-principles_2018} and experimentally.~\cite{miron_current-driven_2010,miron_perpendicular_2011,garello_symmetry_2013}

A prerequisite for the emergence of the spin-orbit torque effect is the existence of strong spin-orbit coupling~\cite{manchon_theory_2009} in materials. This property is shared by topological insulators~\cite{kane_$z_2$_2005,moore_birth_2010,hasan_colloquium:_2010,hasan_three-dimensional_2011} that are narrow-gap semiconductors in the bulk but conducting in the surface due to metallic surface states. The strong spin-orbit coupling causes these states to be topologically protected against surface distortions and gives them a special spin texture with spin momentum locking, leading to the absence of spin degeneracy. Due to spin momentum locking, electrons with opposite group velocities have opposite spin directions. Consequently, the topological insulators display unique and advantageous properties for spin-transport applications.~\cite{pesin_spintronics_2012,mellnik_spin-transfer_2014,wang_topological_2015,fischer_spin-torque_2016} 

\begin{figure}
	\centering
	\includegraphics[width=1\linewidth]{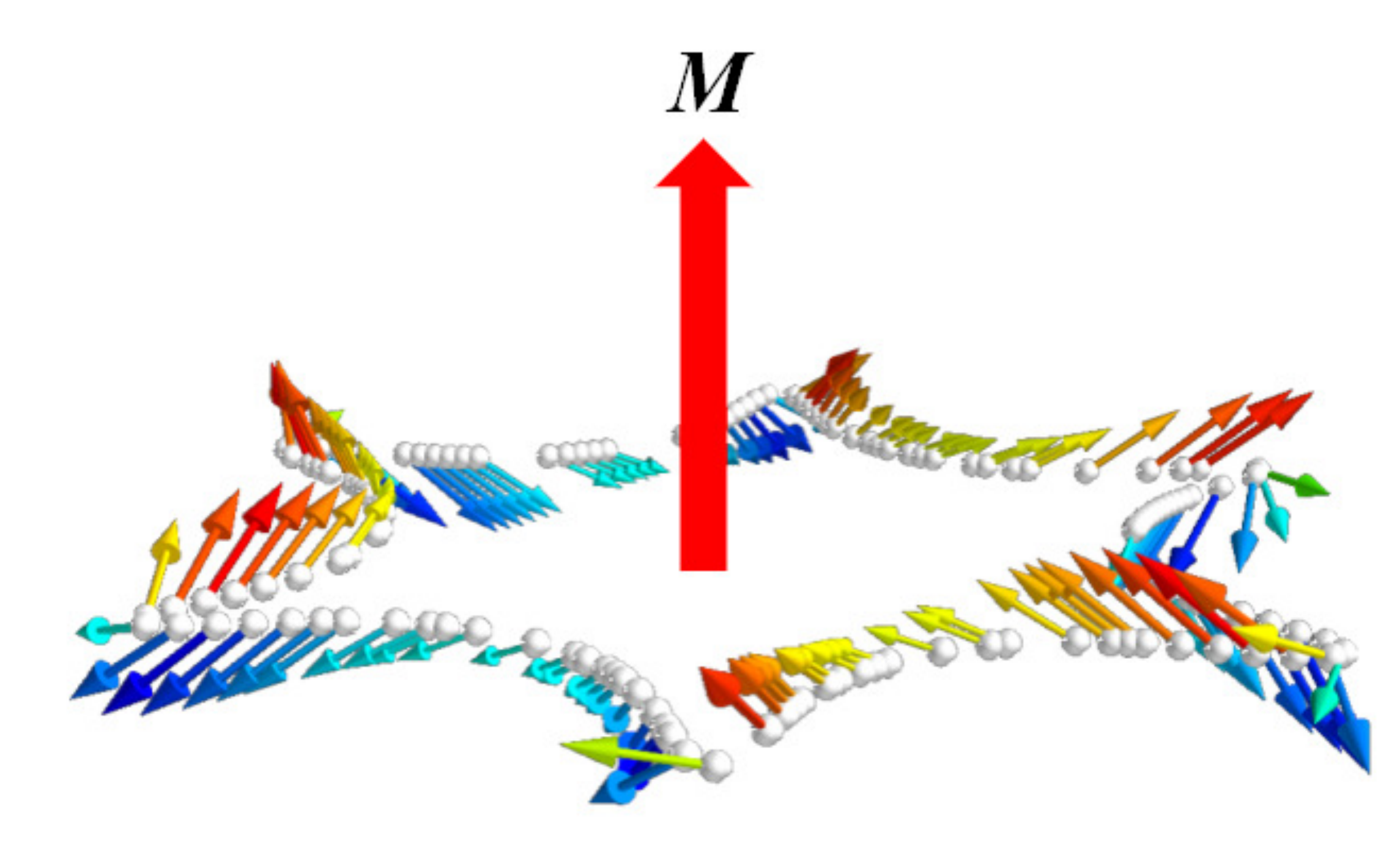}
	\caption{The spin polarization of the Fermi surface states of the topological insulator Bi$ _{2} $Te$ _{3} $ film (side-view). The red arrow in the middle represents the magnetization $\bm{M}$ of the magnetic impurity atom.}
	\label{fig:magnetization}
\end{figure}

The present work focuses on the phenomenon of the SOT on magnetic moments of magnetic transition-metal impurity atoms (Cr, Mn, Fe and Co) embedded in the surface of the topological insulator Bi$_{2}$Te$ _{3}$.~\cite{zhang_topological_2009,hsieh_observation_2009} In these systems the SOT represents the precession of the magnetization of the magnetic impurities in response to an electrical current generated by an electric field in the surface. This precession results from the transfer of spin angular momentum of current-carrying conduction electrons to the magnetic atoms during scattering between surface states of different momentum and consequently different spin polarization. The strong topological insulator Bi$_{2}$Te$_{3}$ is chosen as the substrate, since it is one of the most studied topological insulators. Its simple band structure consists of a single Dirac cone which extends well into the bulk band gap in the vicinity of the $ \Gamma $ point,~\cite{chen_experimental_2009} building a simple hexagonal snowflake-shaped~\cite{fu_hexagonal_2009} Fermi surface. Due to the metallic surface states and the insulating bulk of the topological insulator, all current flows near the surface, where the SOT effect takes place, suggesting optimal efficiency. In addition, the spin polarization $\bm{s}$ of the  Fermi surface states is predominantly in the plane of the Bi$_{2}$Te$ _{3}$ surface and, consequently, perpendicular to the magnetic impurity spin $\bm{M}$ which is taken along the surface normal, maximizing the product $\bm{s}\times\bm{M}$ that governs the torque. The situation is shown in Fig.~\ref{fig:magnetization}. The preferred out-of-plane orientation of the magnetic moment axis has been established by experiments~\cite{sessi_signatures_2014,smann_towards_2018,eelbo_strong_2014} for the Mn/Bi$_{2}$Te$ _{3}$ and Fe/Bi$_{2}$Te$ _{3}$ systems.

The  paper is organized as follows. In Sec.~\ref{matrixelem} we give the formalism for the calculation of the matrix elements of the spin, spin-orbit torque and spin flux operators. In Sec.~\ref{nonequilibrium} we present the approximation, based on the Boltzmann formalism, by which we calculate the non-equilibrium distribution function and the response coefficients in an applied electric field. In Sec. \ref{sc:system} we provide a description of the Bi$_{2}$Te$ _{3}$ system. Sec~\ref{sc:results} includes the results, the correlation of the calculated quantities, and a discussion on the Joule heat production as a function of the  torque. Finally, the main conclusions of this study are summarized in Sec~\ref{conclusions}.

\section{\label{sc:method}Methodology}
The electronic structure of the host and impurity system and the Fermi surface are calculated using the Local Density Approximation (LDA) within Density-Functional Theory (DFT)~\cite{kohn_self-consistent_1965,vosko_accurate_1980} by means of the full-potential relativistic Korringa-Kohn-Rostoker (KKR) Green function method.~\cite{ebert_calculating_2011,long_spin_2014,zimmermann_fermi_2016} The KKR formalism for the calculation of Fermi surfaces, impurity scattering and spin transport using the Boltzmann equation,~\cite{mertig_transport_1999} has been used in the past for problems related to the present study, e.g. for the investigation of the spin Hall~\cite{gradhand_extrinsic_2010,long_spin_2014} and the spin Nerst effect.~\cite{tauber_spin_2013} For this study, the development of the formalism is based on the work of G\'eranton et al.,~\cite{geranton_spin-orbit_2016} who studied the spin-orbit torque effect on the atoms of the magnetic host system FePt. We extend this methodology to the spin orbit-torque effect on the impurity atoms. In particular, we perform the calculations using the impurity scattering wavefunctions instead of the host Bloch wavefunctions. In addition, the multiple scattering of electrons off impurities is included in our study.

\subsection{\label{matrixelem}Spin accumulation, spin-orbit torque and spin flux in the KKR representation}
At first, we calculate the states of the host system $ \psi_{\bm{k}} $ on the Fermi surface, which obey the Bloch's theorem, making use of the KKR secular equation. In a second step, we solve the scattering problem, due to the existence of the impurities in the surface of Bi$_{2}$Te$ _{3}$. The impurity scattering wavefunctions $ \psi^{\mathrm{imp}}_{\bm{k}}$ are calculated by the Lippmann-Schwinger equation:
\begin{equation}\label{lipmann}
\psi^{\mathrm{imp}}_{\bm{k}}(\bm{r}+\bm{R}_{\mu}) = \psi_{\bm{k}}(\bm{r}) + \int d\bm{r'} G^{\mathrm{imp}}(\bm{r},\bm{r'}) \Delta V (\bm{r'}) \psi_{\bm{k}}(\bm{r'}),
\end{equation}
where $ \bm{R}_{\mu} $ is the center of the $ \mu $-th atomic cell. The effect of the impurities in the crystal is described by the perturbing potential $ \Delta V $, which is defined as the difference between the impurity potential $V^{\mathrm{imp}}$ and the potential of the host system $ V^{\mathrm{host}} $ ($ \Delta V = V^{\mathrm{imp}}-V^{\mathrm{host}} $). The impurity Green function $ G^{\mathrm{imp}} $ is related to the Green function of the host system by the Dyson equation $G^\mathrm{imp}=G^\mathrm{host} + G^\mathrm{host} \Delta V G^\mathrm{imp}$.

The knowledge of the scattering wavefunctions allows us to determine the expectation values of the spin, torque and spin flux. The $ i $-th Cartesian component of the spin expectation value for the scattering state $ \bm{k} $ on the Fermi surface integrated in the volume of the atomic cell $ \Omega_{\mu} $ of the impurity atom $ \mu $ is written as
\begin{equation}\label{expvalsigma}
\expval{\sigma_{i\mu}}_{\bm{k}} = \int_{\Omega_{\mu}} d\bm{r}\ [\psi_{\bm{k}}^{\mathrm{imp}}(\bm{r})]^{\dagger} \sigma_{i} [\psi_{\bm{k}}^{\mathrm{imp}}(\bm{r})].
\end{equation}

The torque operator is defined as the external product of the spin with the magnetic part of the exchange-correlation field~\cite{freimuth_spin-orbit_2014,manchon_current-induced_2019}
\begin{equation}\label{torqoperator}
\bm{\mathcal{T}}(\bm{r}) = - \bm{\sigma} \times \bm{B}^{\mathrm{xc}} (\bm{r}),
\end{equation}
where $ \bm{\sigma} $ represents the vector of Pauli matrices. The spin-polarized part of the exchange corellation potential $ \bm{B}^{\mathrm{xc}} $ (in units of energy), is calculated within the LDA, and it is directed opposite to the local magnetization vector $\bm{M}$.\cite{freimuth_spin-orbit_2014} According to the definition of the torque operator, we derive the expression of the $ i $-th torque expectation value for the scattering state $ \bm{k} $ at the impurity atom $ \mu $:
\begin{align}\label{torqueexpectationvalue}
\expval{\mathcal{T}_{i\mu}}_{\bm{k}} = &-\sum_{pq} \epsilon_{ipq}\nonumber\\ &\int_{\Omega_{\mu}} d \bm{r}\ [\psi_{\bm{k}}^{\mathrm{imp}}(\bm{r})]^{\dagger} \sigma_{p} [\psi_{\bm{k}}^{\mathrm{imp}}(\bm{r})] B_{q}^{\mathrm{xc}}(\bm{r}),
\end{align}
where $ \epsilon_{ipq} $ is the Levi-Civita symbol and the indices $ i $, $ p $, $ q $ take the values $ x $, $ y $, and $ z $.
 
We can determine how much of the spin current that enters the sphere which encloses the impurity atom contributes to the spin-orbit torque and how much is lost to the spin lattice interaction by calculating the spin flux. The spin flux operator~\cite{geranton_spin-orbit_2016} is analogous to the spin current operator, but represents the magnetic moment through the spin current which enters the muffin-tin sphere of the atom $ \mu $. The expectation value of the spin flux operator for a Fermi surface state at the impurity atom $ \mu $ is given by the relation~\cite{freimuth_spin-orbit_2014,wessely_current_2006}
\begin{align}\label{spinfluxexpectation}
\expval{\mathcal{Q}_{i\mu}}_{\bm{k}} =\frac{\mu_{B} \hbar}{2ie} \int_{S_{\mu}} d\bm{S} &\Big[[\psi_{\bm{k}}^{\mathrm{imp}}(\bm{r})]^{\dagger} \sigma_{i} \nabla \psi_{\bm{k}}^{\mathrm{imp}}(\bm{r})\nonumber\\ &- [\nabla \psi_{\bm{k}}^{\mathrm{imp}}(\bm{r})]^{\dagger} \sigma_{i} \psi_{\bm{k}}^{\mathrm{imp}}(\bm{r}) \Big],
\end{align}
where $ \hbar $ is the reduced Planck constant, $ e = -\abs{e} $ is the electron's charge, $ \mu_{B} $ is the Bohr magneton, and the integration takes place on the surface $S_{\mu}$ of the muffin-tin sphere of the atom $\mu$.

It is straightforward to cast the above equations into the full-potential relativistic KKR formalism, taking into account the KKR representation of the Green function and wavefunctions. We refer to Refs.~\onlinecite{zimmermann_ab_2014}, \onlinecite{zimmermann_fermi_2016} and \onlinecite{geranton_spin-orbit_2016} for details.

\subsection{\label{nonequilibrium}Non-Equilibrium state}
The calculations of the spin accumulation, the impurity-driven spin-orbit torque, and the spin flux in the non-equilibrium state, applying an external electric field in the system, are based on Boltzmann formalism. 

Within the semiclassical approach, the distribution function of the non-equilibrium system $ f_{\bm{k}} $ is defined as the sum of the equilibrium Fermi-Dirac distribution function $ f^{0}(E_{\bm{k}}) $ and the deviation of the equilibrium $ g_{\bm{k}} $, $ f_{\bm{k}} = f^{0}(E_{\bm{k}}) + g_{\bm{k}} $.

Having solved the multiple scattering problem in the KKR representation, we compute the vector mean free path $ \mathbf{\Lambda}_{\bm{k}} $, solving the self-consistent linearized Boltzmann equation for nominal impurity concentration of 1 (the treatment for the wished concentration is presented in subsection \ref{multiple_scattering})
\begin{equation}\label{selfconspath}
\mathbf{\Lambda}_{\bm{k}} \vdot \hat{n}_{\bm{\mathcal{E}}} = \ \tau_{\bm{k}} \bigg[\bm{v_{\bm{k}}} \vdot \hat{n}_{\bm{\mathcal{E}}}+ \sum_{\bm{k'}} w_{\bm{k}\bm{k'}} (\mathbf{\Lambda}_{\bm{k'}} \vdot \hat{n}_{\bm{\mathcal{E}}})\bigg],
\end{equation}
where $ \bm{v_{\bm{k}}} $ is the group velocity, and $ \hat{n}_{\bm{\mathcal{E}}} =  \bm{\mathcal{E}}/\abs{\bm{\mathcal{E}}}$ is the direction of the electric field $ \bm{\mathcal{E}} $. The scattering rate $  w_{\bm{k}\bm{k'}} $ is expressed in terms of the $ T $-matrix by Fermi's Golden Rule 
\begin{equation}\label{fermigoldenrule}
w_{\bm{kk'}} = \frac{2\pi}{\hbar} \delta(E(\bm{k}) - E(\bm{k'})) \abs{T_{\bm{k}\bm{k'}}}^{2}.
\end{equation}
The transition matrix $ T $ is given by the relation 
\begin{equation}\label{tmatrix}
T_{\bm{k'}\bm{k}} = \int d\bm{r}  \psi^{\dagger}_{\bm{k'}}(\bm{r}) \Delta V(\bm{r}) \psi^{\mathrm{imp}}_{\bm{k}}(\bm{r}).  
\end{equation}
Furthermore, the relaxation time in Eq.~\eqref{selfconspath} is $\tau_{\bm{k}} = 1/ \sum_{\bm{k'}} w_{\bm{k}\bm{k'}}$. It is important to note that our calculations are not based on the independent scattering approximation, as explained in the following Subsection~\ref{sec:aver}.

Once the vector mean free path has been evaluated, we find the linearized expression with respect to the electric field for the deviation of the equilibrium distribution function $ g_{\bm{k}} $
\begin{equation}\label{path}
g_{\bm{k}}=  -e \pdv{f^{0}(E_{\bm{k}})}{E_{\bm{k}}} \mathbf{\Lambda}_{\bm{k}} \vdot \bm{\mathcal{E}}.
\end{equation}
Having estimated the $ g_{\bm{k}} $, one can proceed with the calculation of the spin-orbit torque that is exerted on impurity atom $ \mu $, which is written by means of the deviation distribution function in Boltzmann formalism as
\begin{equation}\label{torq}
\bm{T}_{\mathrm{\mu}} = \sum_{\bm{k}} g_{\bm{k}} \expval{\bm{\mathcal{T}}_{\mathrm{\mu}}}_{\bm{k}}.
\end{equation}
The torque expectation value is computed by Eq.~\eqref{torqueexpectationvalue}. Replacing the deviation distribution function (Eq.~\eqref{path}) in the above equation we obtain the following Fermi surface (FS) integral for the impurity-driven spin-orbit torque
\begin{equation}\label{torqfs}
\bm{T}_{\mathrm{\mu}} = -\frac{e}{\hbar S_{\mathrm{BZ}}} \int_{\mathrm{FS}} \frac{dk}{ \abs{\bm{v_{\bm{k}}}}} (\expval{\bm{\mathcal{T}}_{\mathrm{\mu}}}_{\bm{k}} \otimes \mathbf{\Lambda}_{\bm{k}}) \vdot \bm{\mathcal{E}},
\end{equation}
where $ S_{\mathrm{BZ}} $ is the Brillouin zone surface.
	
In this study we focus on the linear response of the SOT to an external electric field, which is represented by the torkance tensor $ \bm{t}_{\mathrm{\mu}} $~\cite{freimuth_spin-orbit_2014}
\begin{equation}\label{tork}
\bm{T}_{\mathrm{\mu}} = \bm{t}_{\mathrm{\mu}} \bm{\mathcal{E}}.
\end{equation}
It is easily proved by Eqs.~\eqref{torqfs},~\eqref{tork} that the torkance is computed by the expression
\begin{equation}\label{torkance1}
\bm{t}_{\mathrm{\mu}} = -\frac{e}{\hbar S_{\mathrm{BZ}}} \int_{\mathrm{FS}} \frac{dk}{ \abs{\bm{v_{\bm{k}}}}} \expval{\bm{\mathcal{T}}_{\mathrm{\mu}}}_{\bm{k}} \otimes \mathbf{\Lambda}_{\bm{k}}.
\end{equation}
In a similar way the response coefficient of the spin accumulation $ \bm{\chi}_{\mu} $, and the response coefficient of the spin flux $ \bm{q}_{\mu} $ to the electric field are calculated in Boltzmann formalism by the following integrals
\begin{align}\label{spintensor}
\bm{\chi}_{\mu} &= -\frac{e\mu_{B}}{\hbar S_{\mathrm{BZ}}} \int_{\mathrm{FS}} \frac{dk}{ \abs{\bm{v_{\bm{k}}}}} \expval{\bm{\sigma}_{\mu}}_{\bm{k}} \otimes \mathbf{\Lambda}_{\bm{k}},\\
\label{spinfluxtensor}
\bm{q}_{\mu} &=\ \ \frac{e}{\hbar S_{\mathrm{BZ}}} \int_{\mathrm{FS}} \frac{dk}{ \abs{\bm{v_{\bm{k}}}}} \expval{\bm{\mathcal{Q}}_{\mathrm{\mu}}}_{\bm{k}} \otimes \mathbf{\Lambda}_{\bm{k}}.
\end{align}
On the above Eqs.~\eqref{spintensor},~\eqref{spinfluxtensor}, the expectation values of the spin accumulation $ \expval{\bm{\sigma}_{\mu}}_{\bm{k}} $ and the spin flux $ \expval{\bm{\mathcal{Q}}_{\mathrm{\mu}}}_{\bm{k}} $ are determined by Eqs.~\eqref{expvalsigma},~\eqref{spinfluxexpectation}, respectively.

In addition, we can find the current density $ j $, which is calculated by means of the distribution function. According to Ohm's law, the knowledge of the current density allows us to compute the conductivity tensor $ \sigma_{ij} $, as it is readily obtained as the prefactor to the electric field. Thus, the conductivity tensor is given by the equation
\begin{equation}\label{conductivitytensor}
\sigma_{ij} = \frac{e^{2}}{4 \pi^{2}} \int_{\mathrm{FS}} \frac{dk}{\hbar \abs{\bm{v_{\bm{k}}}}} (\bm{v_{\bm{k}}})_{i}\ ( \mathbf{\Lambda}_{\bm{k}})_{j}.
\end{equation}
After computing the diagonal elements of the conductivity tensor, we can estimate the resistivity $ \bm{\rho} $, which is defined as $ \bm{\rho}=\bm{\sigma}^{-1} $.

\subsection{\label{multiple_scattering}Multiple scattering and averaging over configurations\label{sec:aver}}
The Fermi wavevector in Bi$_2$Te$_3$ is of the order of $k_F\approx0.2$\AA$^{-1}$, which gives an estimated Fermi wavelength of $\lambda_F=2\pi/k_F\approx50\text{\AA}\approx12 a_{\rm NN}$, where $a_{\rm NN}=4.38$\AA\ is the nearest-neighbor distance in the surface. At the surface concentrations of 2\% and 5\%, that we wish to study, the average distance between impurities is of the order of $7a_{\rm NN}$ and $4.5a_{\rm NN}$. Clearly, many defects will be present within a radius of one wavelength around the impurity. Therefore, the approximation of independent impurity scattering, that is conventionally used in the Boltzmann equation, becomes questionable. In other words, the scattering rate $w_{\bm{kk'}}$ cannot be approximated by the rate of a single impurity, scaled by the concentration. 

Hence, we take a different approach, where we explicitly consider a collection of $N_{\rm def}= 51$ defects, randomly placed within a circular disc of a radius of a few $\lambda_F$, while outside the disk we consider boundary conditions of the pristine host (see Fig.\ \ref{fig:struc}c). The radius is adjusted so that the number of defects in the disk corresponds to the concentration. Formally, this collection is treated as a single super-impurity, for which the Green function $G^{\mathrm{imp}}$, the scattering states (\ref{lipmann}) and the transition matrix (\ref{tmatrix}) are calculated. The resulting scattering states and scattering rate include the amplitudes and phases of all multiple scattering events off defects within this radius, summed to all orders. In a second step, the scattering rate is scaled by an appropriate concentration of super-impurities, so that the concentration of defects is matched. Thus, if $c$ is the wished defect concentration, then the disk radius is adjusted to enclose $N_{\rm disk}=N_{\rm def}/c$ surface atoms. If the calculated scattering rate by the super-impurity of $N_{\rm def}$ atoms is $w_{\bm{kk'}}$, then we set a concentration of $x_{\mathrm{imp}} =  c/N_{\rm def}$ in the Boltzmann Eq. (\ref{selfconspath})
\begin{align}\label{boltz2}
(x_{\mathrm{imp}} \mathbf{\Lambda}_{\bm{k}} \vdot \hat{n}_{\bm{\mathcal{E}}}) N_{\mathrm{cr}} =& \ \tau_{\bm{k}} \bigg[\bm{v_{\bm{k}}} \vdot \hat{n}_{\bm{\mathcal{E}}} +\nonumber\\ &N_{\mathrm{cr}} \sum_{\bm{k'}} w_{\bm{k}\bm{k'}} (x_{\mathrm{imp}} \mathbf{\Lambda}_{\bm{k'}} \vdot \hat{n}_{\bm{\mathcal{E}}})\bigg],
\end{align}
The problem of finding the Green function of a system with 51 defects poses no numerical difficulty (see Appendix \ref{sc:app2}).

In a third step, we calculate a number of $N_{\rm conf}=20$ different random defects configurations, but always fixing one defect at the center of the disk. We consider this central defect as the most representative of the situation of a homogeneously doped surface. In the results we show the torque acting on the moment of the central defect only.

\section{\label{sc:system}Studied system}

\begin{figure}
	\centering
	\includegraphics[width=0.8\textwidth]{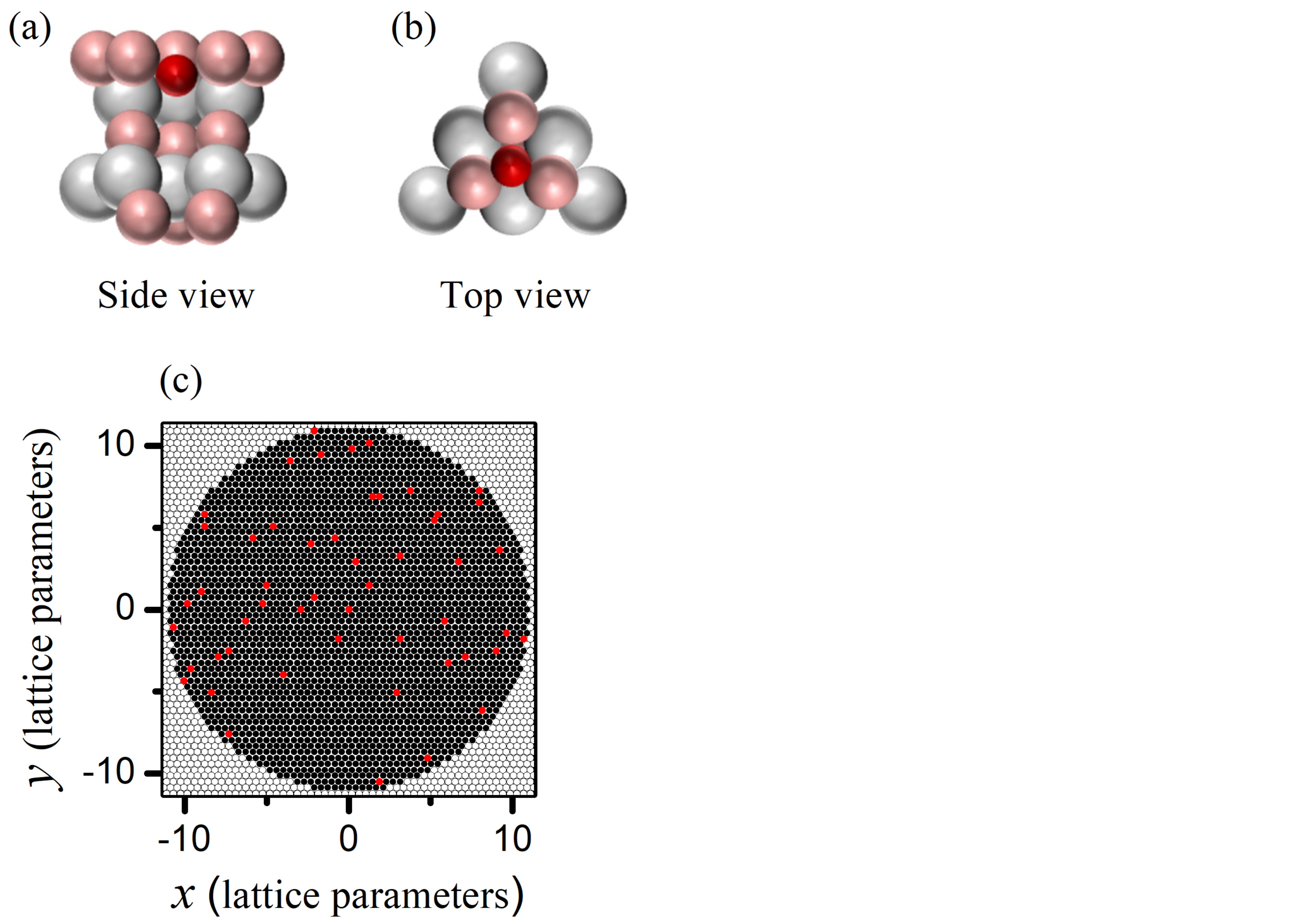}
	\caption{(a) Illustration of the first quintuple layer of Bi$ _{2} $Te$_{3} $ in side-view and (b) top-view. The pink (medium-sized) and the gray (large) spheres represent the Te and Bi atoms, respectively. The magnetic impurity atom is depicted in red (small sized sphere). The impurity shows an inward relaxation with respect to the surface Te layer with vertical distance of 0.9\AA, as has been found for Fe impurities by Eelbo et al.\cite{eelbo_strong_2014}. (c) Schematic representation of the random positions of the defects on the surface in one of the random configurations. The filled red (gray-colored in gray-scale) circles represent the magnetic transition metal defects. The filled black and empty circles depict unoccupied surface impurity sites (threefold hollow positions with fcc stacking with respect to the surface layer), respectively inside and outside the disk in which the 51 impurities are embedded.}
	\label{fig:struc}
\end{figure}

We investigate the surface of Bi$ _{2} $Te$ _{3} $~\cite{zhang_first-principles_2010} doped with magnetic transition-metal impurities. We study the four different defects systems Cr/Bi$ _{2} $Te$ _{3} $, Mn/Bi$ _{2} $Te$ _{3} $, Fe/Bi$ _{2} $Te$ _{3} $, and Co/Bi$ _{2} $Te$ _{3} $. The surface of Bi$ _{2} $Te$ _{3}$, i.e. the structure of the host system, is simulated by a thick film of 6 quintuple layers of Bi$ _{2} $Te$ _{3} $ including 9 vacuum layers on top and bottom to ensure a proper embedding into the vacuum. The impurity atoms are embedded in the interstitial position between the first Te and Bi layer, in fcc hollow site according to the experiments,~\cite{eelbo_strong_2014} as it is shown in Fig.~\ref{fig:struc}(a,b), where the position of the defect in the first quintuple layer is shown from a side view and in a top view, respectively. In particular, the impurity position layer is shifted inward by 0.9\AA\ with respect to first Te layer. For sure, in experiment, the exact position can change for different impurity types. This should have no qualitative consequences on our conclusions, which are related primarily to the simple form of the spin scattering of the Fermi-surface states, as we elaborate in the following sections. 

In the context given in  Sec.\ \ref{sec:aver}, we consider two concentrations: 2\%, corresponding to $N_{\rm def}=51$ defects, randomly placed within a disk of $N_{\rm disk}=2539$ positions, and 5\%, corresponding to $N_{\rm def}=51$ defects within a disk of $N_{\rm disk}=1027$ positions. A statistical averaging is achieved by considering $N_{\rm conf}=20$ different random configurations. For comparison with the conventional Boltzmann formalism, we also calculate results using the scattering rate from a single defect (neglecting multiple scattering).

We take the defect magnetic moments to be perpendicular to the surface, in accordance to findings ~\cite{sessi_signatures_2014,smann_towards_2018,eelbo_strong_2014} for the Mn/Bi$_{2}$Te$ _{3}$ and Fe/Bi$_{2}$Te$ _{3}$ systems. Furthermore, we assume a ferromagnetic alignment of the magnetic defects, as has been observed experimentally at 2\% concentration for Mn defects and at $>3$\% for Co defects.~\cite{smann_towards_2018} Of course, the aforementioned assumptions have not been found for all considered defect types at all concentrations (e.g., for Co, antiferromagnetic interactions appear at 2\% concentration~\cite{smann_towards_2018}). Extending the assumptions of out-of-plane orientation and ferromagnetism to all cases should be considered a numerical experiment. We know from previous studies~\cite{smann_towards_2018} that ferromagnetic interactions can be engineered by appropriate doping that shifts the Fermi level of the system and could conceivably be achieved in all four types of defects. Analogous engineering is conceivable for the magnetic anisotropy. In addition, by treating all types of defects on the same footing, we gain understanding of the chemical trends of the SOT mechanism.

\section{\label{sc:results}Results and Discussion}

\begin{figure*}
	\centering
	\includegraphics[width=1\textwidth]{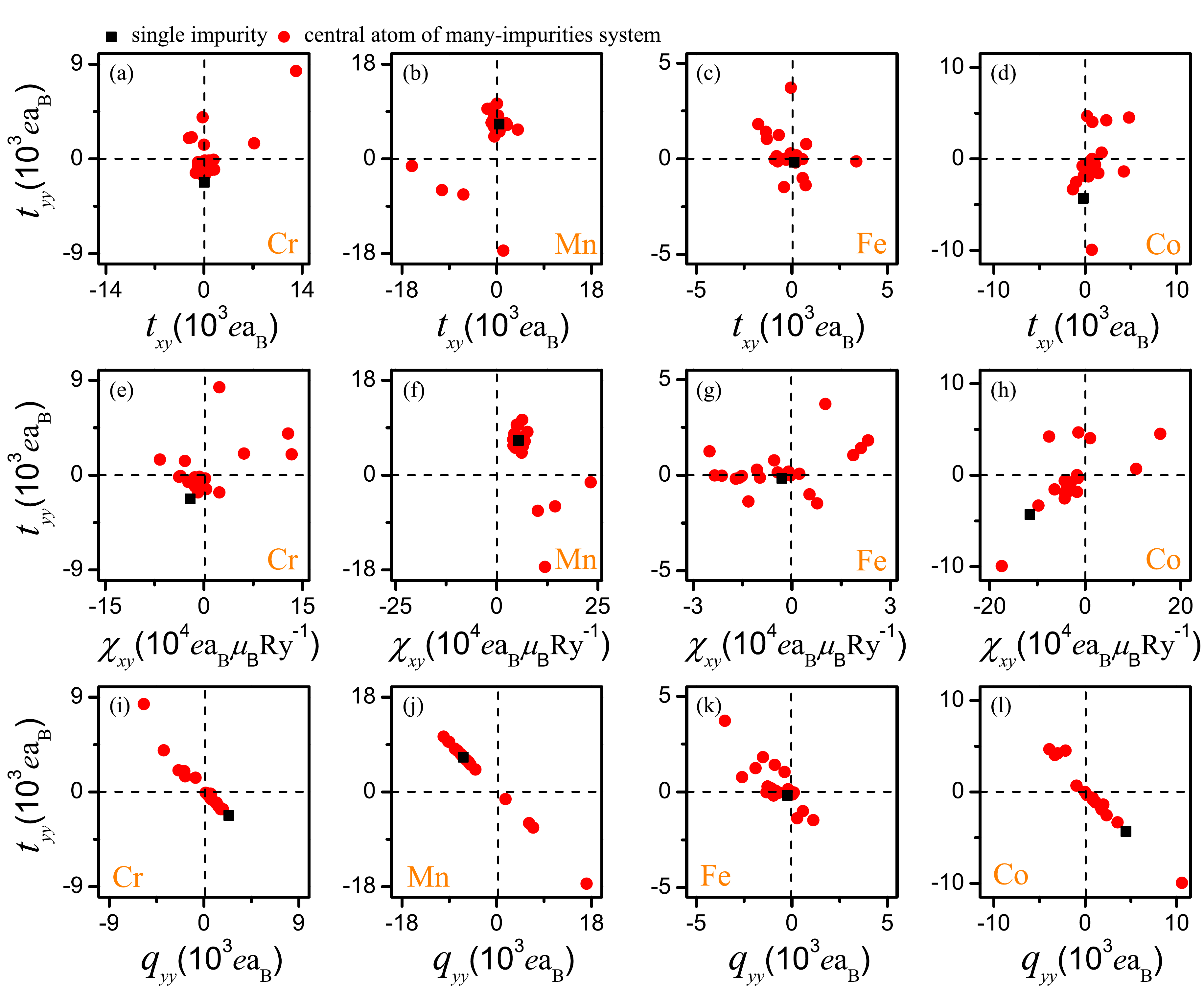}
	\caption{The $ y $ component of the torkance $ t_{yy}$ as a function of the $ x $ component of the torkance $t_{xy}$ (a-d), the torkance $t_{yy}$ as a function of the response coefficient of the spin accumulation $\chi_{xy} $ (e-h), the torkance $ t_{yy}$ as a function of the response coefficient of the spin flux $ q_{yy} $ (i-l), on the central Cr, Mn, Fe, and Co impurity atom in the presence of 1 defect (squares) and 2\% defects concentration for 20 different distributions (circles), embedded in Bi$ _{2}$Te$ _{3} $ surface. The results are scaled to a 2\% concentration of defects. The electric field is taken in $ y $ direction. The torkance is given in units of $e\mathrm{a}_{\mathrm{B}}=9 \times 10^{-5}\mu_{\mathrm{B}}$T/(V/cm). The spin accumulation is given in units of $e\mathrm{a}_{\mathrm{B}}\mu_{\mathrm{B}}\mathrm{Ry}^{-1}=3 \times 10^{-10}\mu_{\mathrm{B}}$/(V/cm).}
	\label{fig:ty_tx}
\end{figure*}

\subsection{Computational details}
The density functional calculations for the electronic structure of the Bi$ _{2} $Te$_{3} $ film were carried out with the J\"ulich full potential relativistic KKR code.~\cite{noauthor_jukkr_nodate} For the computation of the Green functions a finite angular momentum cutoff of $ l_{\mathrm{max}}=3 $ was used.

The self-consistent potential of the impurity atoms was computed using the J\"ulich KKR impurity-embedding code KKRimp~\cite{bauer_development_2014} in a cluster including the 14 nearest neighboring sites of the defect, which is sufficient for the correct charge screening due to the metallic surface states. The impurity-atom potential is then placed in the respective 51 random impurity positions. This approximation saves computational time compared to a fully self-consistent calculation of the system of 51 impurities together. Tests have shown that the approximation is adequate for the description of the potential, if the impurities occupy farther than nearest-neighbour positions, which holds for the great majority of cases at low concentration. The multiply scattered wavefunction and the scattering rate are calculated in this way.

Fermi surface calculations as well as Boltzmann transport computations were performed with the J\"ulich PKKRcode.~\cite{zimmermann_fermi_2016} For the calculations of the response coefficients we used 78 $ k $-points in the full Fermi surface of Bi$_2$Te$_3$, which is adequate, since the FS consists of only a single closed loop near the center of the SBZ.

\subsection{\label{torkance}Response coefficients to the electric field} 
In this Section we present and discuss the results of our study. Applying the Boltzmann formalism outlined in Section~\ref{sc:method}, we performed calculations for the response coefficients of the spin accumulation $ \bm{\chi} $, the spin-orbit torque $ \bm{t} $ and the spin flux $ \bm{q} $ in the electric field, that are exerted on the magnetic moment of the impurity atoms embedded in Bi$ _{2} $Te$_{3} $ surface. 

We present the results of the tensor components in response to the electric field in $ y $ direction $ \mathcal{E}_{y} $, which respects the reflection symmetry over the $ y-z $ plane in the host structure. Obviously, the $ z $ component of the torkance is zero, since the moments point along the z axis.

At first, the response coefficient of the spin-orbit torque (torkance) to the electric field $ \mathcal{E}_{y} $ is shown in Fig.~\ref{fig:ty_tx}. Comparing the results of the single defect system with the corresponding results for the central atom of the many impurities system for the different configurations, which is shown in Figs.~\ref{fig:ty_tx}(a-d), we find that the single impurity system is not representative in general. This is anticipated, as the independent scattering approximation is not valid in this system, in other case the average torkance over the many impurities systems would correspond to the single impurity system (The reader can find the detailed analysis in Appendix~\ref{sc:appendix}). Instead, we observe that the torkance presents a spread for all different types of impurities systems. We also find that the largest value of the torkance is exerted on the Mn moment. 

As it is shown in Figs.~\ref{fig:ty_tx}(e-h), where the torkance versus the response coefficient of the spin accumulation is plotted, there is no linear correlation between the spin of the conduction electrons and the spin-orbit torque, as one might except from simple models. This absence of linear correlation is due to the atom size, as the torkance is calculated by a convolution involving one integral which includes the external product (see Eq. \eqref{torqueexpectationvalue}), and it is not a product of the total spin and magnetic field.

\begin{figure}
	\centering
	\includegraphics[width=1\linewidth]{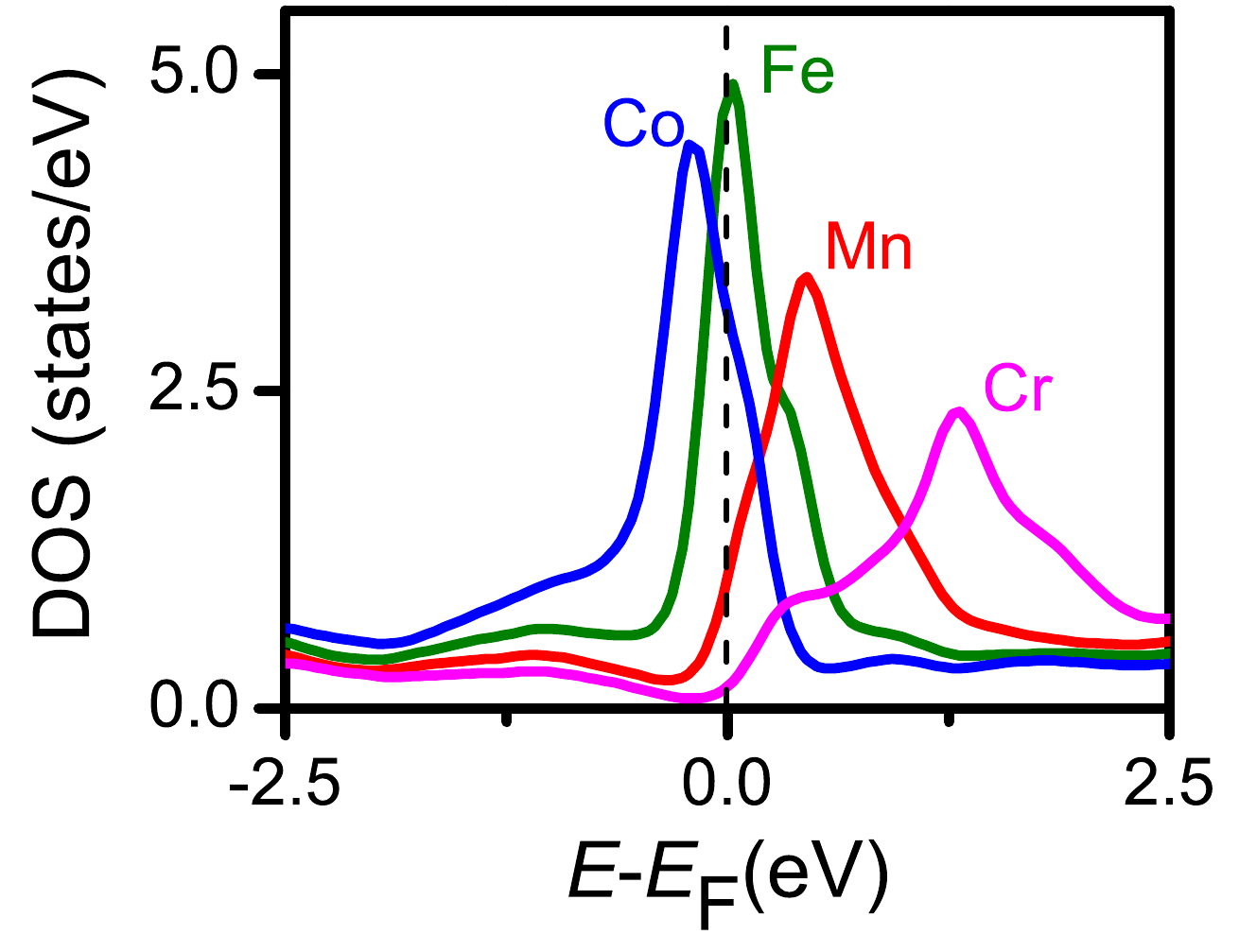}
	\caption{The local spin-resolved density of states (DOS) of the Cr, Mn, Fe and Co impurity atoms.}
	\label{fig:dos}
\end{figure}

In the following, the response coefficient of the spin flux is investigated. In Figs.~\ref{fig:ty_tx}(i,j,l) we observe that the torque has almost a linear dependence on the spin flux for the system of Bi$ _{2} $Te$ _{3} $ doped with Cr, Mn, and Co impurities. This demonstrates that the SOT exerted on the impurity moment is essentially mediated by spin currents in these systems, while the spin-lattice contribution due to the spin-orbit coupling in the impurity atomic sphere is negligible. On the contrary, the spin-lattice interaction is significant on the Fe impurities system, as it is shown in Fig.~\ref{fig:ty_tx}(k) there is still a correlation between the SOT and the spin flux, but not as strong as the other impurities systems. The latter indicates that in the Fe/Bi$ _{2} $Te$ _{3} $ system a part of the current contributes to the spin precession of the Fe impurity, while the rest is lost to the spin-lattice interaction. By the density of states (DOS) of the impurity atoms (Fig.~\ref{fig:dos}), it is observed that the Fe impurity presents a resonance exactly on the Fermi level, whereas the resonance of the other impurity atoms (Co, Mn, Cr) is somewhat shifted with respect to the Fermi energy. Therefore, a longer delay time~\cite{bohm_quantum_1993} of the scattered conduction electron in the Fe system is expected, i.e. the electrons interact a longer time with the spin-orbit field of the nucleus.~\cite{heers_effect_2011-1} As a result, there is a strong interaction of the spin with the lattice in this system.

\begin{figure}
	\centering
	\includegraphics[width=1\linewidth]{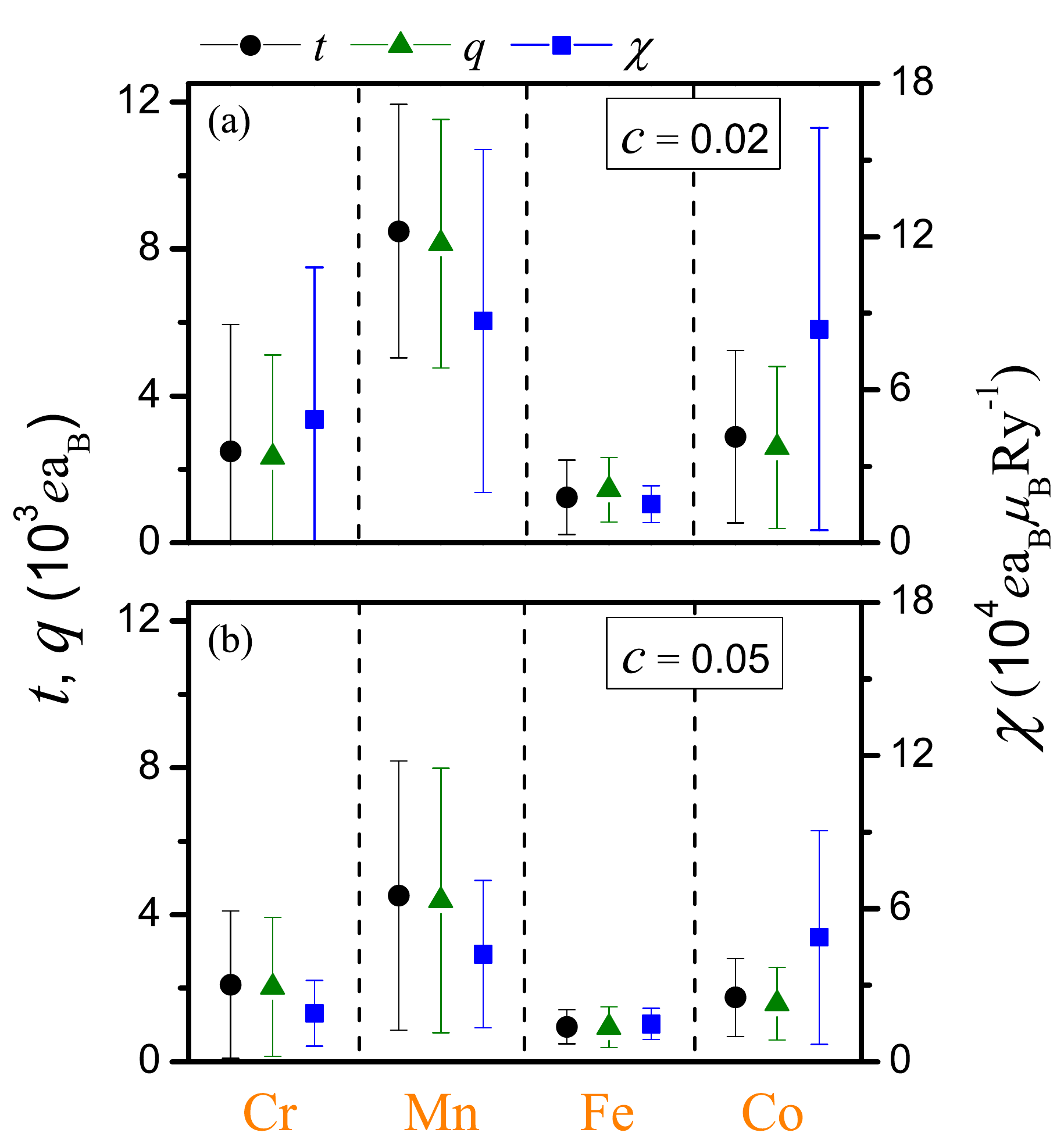}
	\caption{(a) The average torkance $ t $ (circles), spin flux $ q $ (triangles) and spin accumulation $ \chi $ (squares) on the central atom in the presence of 2$ \% $ (a) and 5$ \% $ (b) concentration of Cr, Mn, Fe, and Co impurities embedded in the Bi$ _{2} $Te$ _{3} $ surface. The values are averaged over the 20 different configurations and the error bars indicate the standard deviation of the values.}
	\label{fig:mean_values}
\end{figure}

Next, we compare the results of the systems with 2\% and 5\% defects concentration, in order to find how the impurities concentration affects the spin, spin-orbit torque, and spin flux. The absolute response coefficients of the averaged spin-orbit torque $t=\sqrt{(t_{xy})^{2}+(t_{yy})^{2}} $, the spin flux $q=\sqrt{(q_{xy})^{2}+(q_{yy})^{2}} $, and the spin accumulation $\chi=\sqrt{(\chi_{xy})^{2}+(\chi_{yy})^{2}} $ in an applied electric field $ \mathcal{E}_{y} $ are presented for the two different defect concentrations in Fig.~\ref{fig:mean_values}. Comparing Figs.~\ref{fig:mean_values}(a) and~\ref{fig:mean_values}(b), we find that the magnitude of the torkance, the spin flux and the spin accumulation is greater in the case of lower concentration for all types of impurities. This observation is consistent with the fact that a lower concentration leads to a less perturbed topological surface state. This case is closer to the ideal situation, where the electron states incident on the defects have their spin in-plane, perpendicular to the defect magnetization, and produce maximal torque. The Mn/Bi$_{2}$Te$_{3}$ system displays the largest torkance at both concentrations, in line with the results of Fig.~\ref{fig:ty_tx}. The lowest torkance is seen in the Fe/Bi$_{2}$Te$_{3}$ system, for which we expect the strongest resonant scattering. 

For practical applications we are also interested in the time needed for a reversal of the impurity moment direction. We can estimate this by means of the angular rotation velocity per unit electric field which normalises the torkance to the impurity moment modulus $m_{\mathrm{at}}$:
\begin{equation}\label{rotvel}
\omega = \frac{1}{\mathcal{E}}{\Dot{\theta}} =\frac{2\mu_{B}}{\hbar m_{\mathrm{at}}} t.
\end{equation}

Moreover, according to Eq.~\eqref{conductivitytensor}, we can compute the longitudinal resistivity ($ \rho_{yy} = \sigma_{yy}^{-1}$) of the impurity atoms. By the knowledge of the torkance and the resistivity, the torque for a given current density $j_{y}$ can be derived. We define the linear-response coefficient
\begin{equation}\label{torkres}
\tilde{t}=\frac{T}{j_{y}} = t \vdot \rho_{yy}.
\end{equation}
Knowledge of this quantity serves two purposes. First, it promotes the viewpoint of the torque resulting as a response to the current, instead of the electric field. This picture is convenient especially in magnetic-impurity systems: we have the spin of the current-carrying electronic states of the host, on the one hand, and the electronic and magnetic structure of the impurity, on the other hand. The interaction of the two, due to spin scattering, produces the torque. The electric field does not enter the above picture, even though in reality it is the cause of the current.

The second purpose of introducing $\tilde{t}$, is that its product with the torkance, $(\tilde{t}\,t)$, is related to the Joule heat produced per unit time and volume, $\dot{Q}$, in order to achieve a given torque value $T$: 
\begin{equation}\label{jouleheat}
   \dot{Q}=\rho_{yy}\,j_y^2 = \frac{T^2}{\tilde{t}\,t}.
\end{equation}
What we calculate here is actually a lower bound to the Joule heat, assuming that the magnetic-impurity scattering is the dominant source of resistivity.

\begin{figure}
	\centering
	\includegraphics[width=1\linewidth]{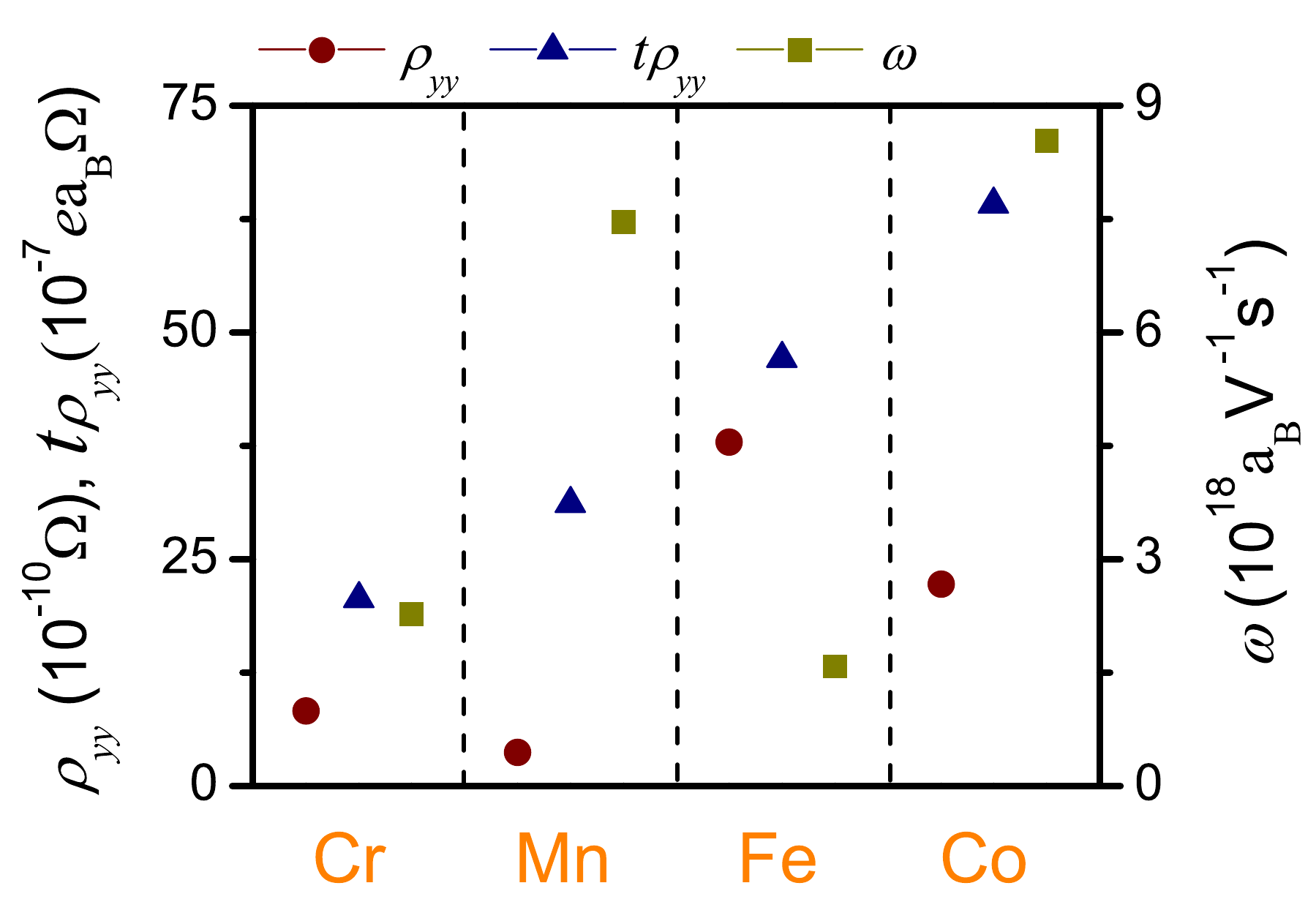}
	\caption{The resistivity $ \rho_{yy} $ (circles), the torkance on the central impurity atom multiplied by the resistivity $\tilde{t}=t\rho_{yy}$ (Eq.\ \ref{torkres}) (triangles), and the angular rotation velocity $\omega$ per unit electric field (Eq.\ \ref{rotvel}) (squares), averaged over the 20 different configurations, in the presence of 2\% defects concentration in the (Cr, Mn, Fe, Co)/Bi$ _{2} $Te$ _{3} $ systems. The electric field is taken in $ y $ direction.}
	\label{fig:conductivity}
\end{figure}

The computed magnetic moments of the defects are presented in Table \ref{magneticmoments}. The results of the resistivity, the ratio of the SOT to the current density and the rotation velocity for Cr, Mn, Fe and Co impurity atoms are depicted in Fig. \ref{fig:conductivity}. One can easily observe that the Mn/Bi$ _{2} $Te$ _{3} $ system presents the lowest resistivity, a large spin-orbit torque for a given current, and large rotation velocity. As a consequence, this system is optimal for applications. Although a large torque for a given current is calculated in the Fe/Bi$ _{2} $Te$ _{3} $ system, this system presents the largest resistivity due to the resonant scattering of the Fe atoms, rendering it less optimal for applications.

The factor $\langle (\tilde{t}\,t)^{-1} \rangle$, averaged over the 20 configurations, is presented in Table~\ref{jouleheatfactor}. We find that the Joule heat for a given spin-orbit torque is much smaller in Mn/Bi$ _{2} $Te$ _{3} $ than the other impurity types systems, that results in fastest and energetically most efficient switching, i.e. has the lowest resistivity and the Joule heat production.

\begin{table}
	\begin{tabular}{l c c}
		Impurity type   \ \ \ \ \ \  & \ \ $ m_\mathrm{at} $ ($ \mu_{B} $) \ \ & \ \ $\langle (\tilde{t}\,t)^{-1}\rangle $ (S/($e^{2}$a$ _{\mathrm{B}}^{2} $)) \\
		\hline
		Cr & 3.331 & 2497.407\\
		Mn & 3.456 & 51.207\\
		Fe & 2.382 & 1160.321\\
		Co & 1.027 & 175.831\\	
		\hline
		\hline
	\end{tabular}
	\caption{The computed spin magnetic moments $ m_\mathrm{at} $ of the magnetic defects and their Joule heat factor $\langle (\tilde{t}\,t)^{-1}\rangle$.}
	\label{magneticmoments}
	\label{jouleheatfactor}
\end{table}

\section{\label{conclusions}Conclusions}
In summary, we have applied the full-potential relativistic KKR Green function method and the Boltzmann formalism to perform calculations of the response coefficients of the spin, spin-orbit torque and spin flux on magnetization of magnetic impurity atoms (Cr, Mn, Fe, Co) embedded in the surface of the topological insulator Bi$ _{2} $Te$ _{3} $. The methodology we employed, takes into account the multiple scattering off impurities. 

We found a strong spin-orbit torque effect in Bi$ _{2} $Te$ _{3} $ doped with magnetic transition-metal atoms. Our findings validate that topological insulators with a simple bandstructure are favorable materials for application of the spin-orbit torque effect. The main reason for this enhanced effect is due to its special characteristics, i.e. the localized surface states and the perpendicular spin polarization of the surface states with respect to the magnetization of the defects. Finally, we predict that the Mn/Bi$ _{2} $Te$ _{3} $ system is the most promising. Our results show that this system presents the lowest resistivity, large spin-orbit torque for a given current, and a large rotation velocity for a specific electric field. In addition, the magnetic moment of the Mn impurity atom is the largest of the four impurity types systems and the Joule heat factor was calculated at least one order of magnitude lower compared to the other impurity types that we considered. Another important characteristic is that it has been shown theoretically and experimentally\cite{smann_towards_2018} that the Mn impurity atoms embedded in Bi$ _{2} $Te$ _{3} $ surface present ferromagnetic behavior in the case of 2\% concentration, which adds confidence on the prospects for application of this system.

\acknowledgments{We thank Y. Mokrousov for fruitful discussions. The research work was supported by the Hellenic Foundation for Research and Innovation (HFRI) under the HFRI PhD Fellowship grant (Fellowship Number: 1314). This work was supported by computational time granted from the Greek Research \& Technology Network (GRNET) in the National HPC facility-ARIS-under projects ID pr00504-TopMag and pr007039-TopMagX. P.R. and P.M. acknowledge funding from the Priority Programme SPP-1666 Topological Insulators of the Deutsche Forschungsgemeinschaft (DFG) (project MA4637/3-1) and support by the Deutsche Forschungsgemeinschaft (DFG, German Research Foundation) under Germany's Excellence Strategy – Cluster of Excellence Matter and Light for Quantum Computing (ML4Q) EXC 2004/1 – 390534769.}

\appendix

\section{Numerical considerations on the Green function of the system with defects\label{sc:app2}}
The Green function of the impurity system is calculated in the KKR method by means of the algebraic Dyson equation
\begin{eqnarray*}\label{eq:aldyson}
	G^{\rm imp,nn'}_{LL'} &=& G^{nn''}_{LL'} + \\
	&& \sum_{n''L''L'''}G^{nn'}_{LL''} (t^{\rm imp,n''}_{L''L'''}-t^{n''}_{L''L'''})
	G^{\rm imp,n''n'}_{L'''L'}
\end{eqnarray*}
where $n$, $n'$, $n''$ are atom-site-indices and $L,L',\ldots$ are indices combining the angular-momentum and spin of an atom at a site. $t(E)$ and $t^{\rm imp}(E)$ are the $T$-matrices of the host and impurity atoms, respectively. $G^{\rm imp}(E)$ is the unknown matrix for the Green function of the system with impurities and $G(E)$ is the known matrix of the host system. This linear set of equations has a dimension proportional to the number of sites for which $t^{\rm imp,n''}_{L''L'''}(E)\neq t^{n''}_{L''L'''}(E)$, i.e., to the number of defects. In this way the problem at hand becomes numerically tractable, since the number of defects (51) that we place in the disk results in a linear system of dimension $N_{\rm def}\times 2(l_{\rm max}+1)^2=1632$, where the number of spin and angular-momentum components at a cutoff of $l_{\rm max}=3$ has been accounted for [$2(l_{\rm max}+1)^2=32$].

\section{\label{sc:appendix}Independent scattering approximation}
To investigate the independent scattering approximation we analyze the inverse relaxation time $\tau_{\bm{k}} = 1/ \sum_{\bm{k'}} w_{\bm{k}\bm{k'}}$ which represents the scattering rate. The approach of independent scattering behind the Boltzmann equation is critically examined, by comparing the calculated scattering rate off single impurity versus multiple defects system. In Table~\ref{tau} the ratio of the average scattering rate of the many defects system for the different configurations to the scattering rate of the single defect system is presented. There is no linear scaling of the scattering rate with the number of impurities in the system. This is consistent with the observation that the Fermi wavelength is longer than the average distance between the impurities. Loosely speaking, after a scattering event of a wavepacket off of a defect, there is not enough space for a new wavepacket to be formed, before it is scattered from another impurity. 

\begin{table}
	\begin{tabular}{l c}
		Impurity type   \ \ \ \ \ \  & \ \ $\tau^{-1}_{51\mathrm{imp}}$/$\tau^{-1}_{1\mathrm{imp}}$  \\
		\hline
		Cr & 141.342 \\
		Mn & 248.066 \\
		Fe &  68.546 \\
		Co & 260.622 \\	
		\hline
		\hline
	\end{tabular}
	\caption{The average scattering rate of the many impurities systems for the 20 different configurations $ \tau^{-1}_{51\mathrm{imp}} $ divided by the scattering rate of the single impurity system $ \tau^{-1}_{1\mathrm{imp}} $, in the presence of 2\% defects concentration in the (Cr, Mn, Fe, Co)/Bi$ _{2} $Te$ _{3} $ systems.}
	\label{tau}
\end{table}

\bibliography{MyLibrary}
\end{document}